# Imaging of Single La Vacancies in LaMnO$_3$

Jie Feng[1], Alexander V. Kvit[1], Chenyu Zhang[1], Jason Hoffman[2], Anand Bhattacharya[2], Dane Morgan[1*], Paul M. Voyles[1*]

[1.] Department of Materials Science and Engineering, University of Wisconsin-Madison, Madison, Wisconsin, 53706, USA

[2.] Materials Science Division, Argonne National Laboratory, Argonne, Illinois 60439, USA

* ddmorgan@wisc.edu, paul.voyles@wisc.edu

## Abstract

We report an approach for three-dimensional imaging of single vacancies using high precision quantitative high-angle annular dark-field Z-contrast scanning transmission electron microscopy (STEM). Vacancies are identified by both the reduction in scattered intensity created by the missing atom and the distortion of the surrounding atom positions. Vacancy positions are determined laterally to a unique lattice site in the image and in depth to within one of two lattice sites by dynamical diffraction effects. 35 single La vacancies are identified in images of a LaMnO$_3$ thin film sample. The vacancies are randomly distributed in depth and correspond to a La vacancy concentration of 0.79%, which is consistent with the level of control of cation stoichiometry within our synthesis process (~1%) and with the equilibrium concentration of La vacancies under the film growth conditions. This work demonstrates an approach to characterizing low concentrations of vacancies with high spatial resolution.

**Introduction**

Point defects, consisting of one foreign, misplaced, or missing atom, are one of the basic classes of deviations from perfect crystal structures in materials. Control of point defects is essential to materials performance, and engineering of point defect populations and distributions with nanometer-scale resolution is becoming as an important tool in designing materials with new functionality[1,2]. High-resolution electron microscopy is emerging as a general tool to characterize point defects with single-defect sensitivity and sub-unit cell spatial resolution in all three dimensions. Detectable defects include substitutional impurities[3–5], interstitial impurities[6], self-interstitials[7], and impurity-containing defect complexes[8,9]. Electron microscopy offers direct information about atomic positions, unmediated by an electronic or vibrational signature of the defect, and atomic resolution, which has led to discovery of new point defect configurations[6] and new impurity-defect complexes[8,9]. However, all the defects so far imaged at the single-defect level increase the local electron scattering, often by 50% or more[3,10]. Defects that decrease scattering, particularly vacancies that change the intensity by 10% or less, have been characterized one at a time only very recently,[11] and not in three dimensions.

Here we demonstrate imaging of single La vacancies in $LaMnO_3$ (LMO) at sub-unit cell resolution in all three dimensions. Understanding the behavior of single vacancies in perovskite structures is critical, as these materials are frequently used as ion conductors and catalysts (e.g., in cathode materials in solid oxide fuel cells (SOFCs)), and their performance is intimately tied to their defect chemistry. In SOFC applications, significant effort has been focused on engineering perovskites with high densities of near-surface vacancies[12], so obtaining 3D, atomic scale information of single vacancies is of great importance for optimizing the performance of SOFC.

Our approach is based on high precision high-angle annular dark-field (HAADF) scanning transmission electron microscopy (STEM), which can measure changes in intensity as small as 1% and shifts in atomic column positions with sub-picometer precision[13]. High precision STEM enables us to image both the decrease in scattered intensity from a vacancy and the shifts of the surrounding atoms in towards the vacancy site. Vacancies are localized laterally to a single column of atoms in the STEM image, and localized in depth to one or two atomic layers by dynamical scattering effects[4,5]. Dynamical scattering is accounted for using quantitative matching of experimental images to frozen-phonon multislice simulations[14,15]. Previous characterization of vacancies by electron microscopy (EM) methods are only sensitive to many vacancies at once and have poorer spatial resolution[16–19] or cannot distinguish bulk vacancies to surface vacancies[11] or visualized single vacancies only in 2D materials[20–22]. Our method is able to overcome the disadvantages all at once.

**Results**

**DFT calculations**

We applied density functional theory (DFT) calculations to study the La vacancy structure in LaMnO$_3$. DFT is notoriously difficult on transition metal oxides, producing results that are strongly dependent on the choice of exchange functional or simply at odds with experiment.[23][1,2,3] However, previous work demonstrates that LaMnO$_3$ is accurately modeled by DFT using the GGA+U, including the formation energetics[24], lattice parameter and band gap[25], and, mostly importantly for this study, it has produced vacancy and other defect formation energies that agree well with bulk experimental measurements.[26][27] We have taken special care to simulate accurately the structure of vacancies in LaMnO$_3$ using DFT. The LaMnO$_3$ structure we studied belongs to the *Pnma* space group and has strong octahedral tilt (Fig. SX in supplemental information) since the STEM experimental temperature is below the 750 K Jahn-Teller distortion transition temperature. We used a 360-atom supercell (more than twice the volume of previous defect structure studies[28][23]) to ensure that the results are not affected by the period boundary conditions, and the ion positions were relaxed until the Hellmann-Feynman forces were less than 1 meV/Å, which 10 times more stringent than previous studies[28][23] (see Experimental Section for details).

Fig. 1A shows the DFT calculated La vacancy structure in the supercell along [100] axis with the vacancy in the center La column. The CONTCAR of VASP calculation is available as supplemental information. In 3D around the La vacancy, the separation between a first nearest-neighbor (NN) La and a second NN La, which is marked as $S_1$ in Fig. 1D, is reduced by 20.3 pm, and the separation between two third NN La atoms is reduced by 21.9 pm, marked as $S_2$ in Fig 1D. This lattice contraction is consistent with the experimentally observed decrease in $LaMnO_3$ lattice parameter with increasing cation vacancy content[29]. Furthermore, due to the small bandgap ($LaMnO3$ (*Pnma*) gap is < 0.8 eV[27]) and generally weak localization of electrons in $LaMnO_3$, the La vacancy structure is independent of its charge state (see Supplementary Information Note 5 for details of charged defect calculations).

**Frozen phonon simulations and STEM experiment**

Fig. 1B is an experimental high-precision HAADF STEM image along the $LaMnO_3$ [100] zone axis. We choose [100] as the zone axis for STEM experiments due to the large La atom separation along this direction (5.5 Å) and the La-O columns are will aligned, as shown in Fig. 1A. The specimen thickness is 6.5 nm measured by position averaged convergence beam electron diffraction (PACBED)[30] (see Supplementary Information Note 2 and Fig. S4). PACBED also confirms that the sample has a mistilt <0.5 mrad, so the slight asymmetry in the image most likely arises from parasitic aberrations like two-fold astigmatism or coma. Such aberrations do not change the integrated intensity of the atomic column images[31]. The simulated image from frozen phonon simulation at the same specimen thickness and collection angle is in the white dashed box.

In both experimental and simulated HAADF STEM images, each La-O column was fit to a two-dimensional (2D) Gaussian function to determine the atomic column position and calculate the column intensity[13]. A vacancy decreases the intensity of the column, which we measure using the visibility, $V$, defined as

$$V = \left(1 - \frac{I}{I_{ref}}\right) \times 100\%, \tag{1}$$

where $I$ is the integrated intensity of the studied column. In simulated images, $I_{ref}$ is the integrated intensity of a vacancy-free reference column with the same geometric thickness, and in experimental images, $I_{ref}$ is the mean integrated intensity of the eight surrounding columns of the studied column. A vacancy also causes an inward shift of the surrounding atoms toward the vacancy as shown in Fig. 1A. We measure this effect as a shift of the neighboring La-O columns in the STEM images, defined graphically in Fig. 1D, and quantitatively by

$$\Delta S = S_{ref} - S, \tag{2}$$

where $S$ is the inter-column separation surrounding a particular column and $S_{ref}$ is the same separation surrounding a vacancy-free column at the same geometric thickness.

We applied frozen phonon simulations to a series of $LaMnO_3$ models with a single La vacancy at different depths[14][32] (See Experimental Section for details). Fig. 1C shows how the simulated visibility, $V^s$, from a single vacancy in a La-O column changes as a function of the La vacancy depth. In a 10 nm thick specimen (solid), $V^s$ is 2% – 7%, so the vacancy is detectable ($V$

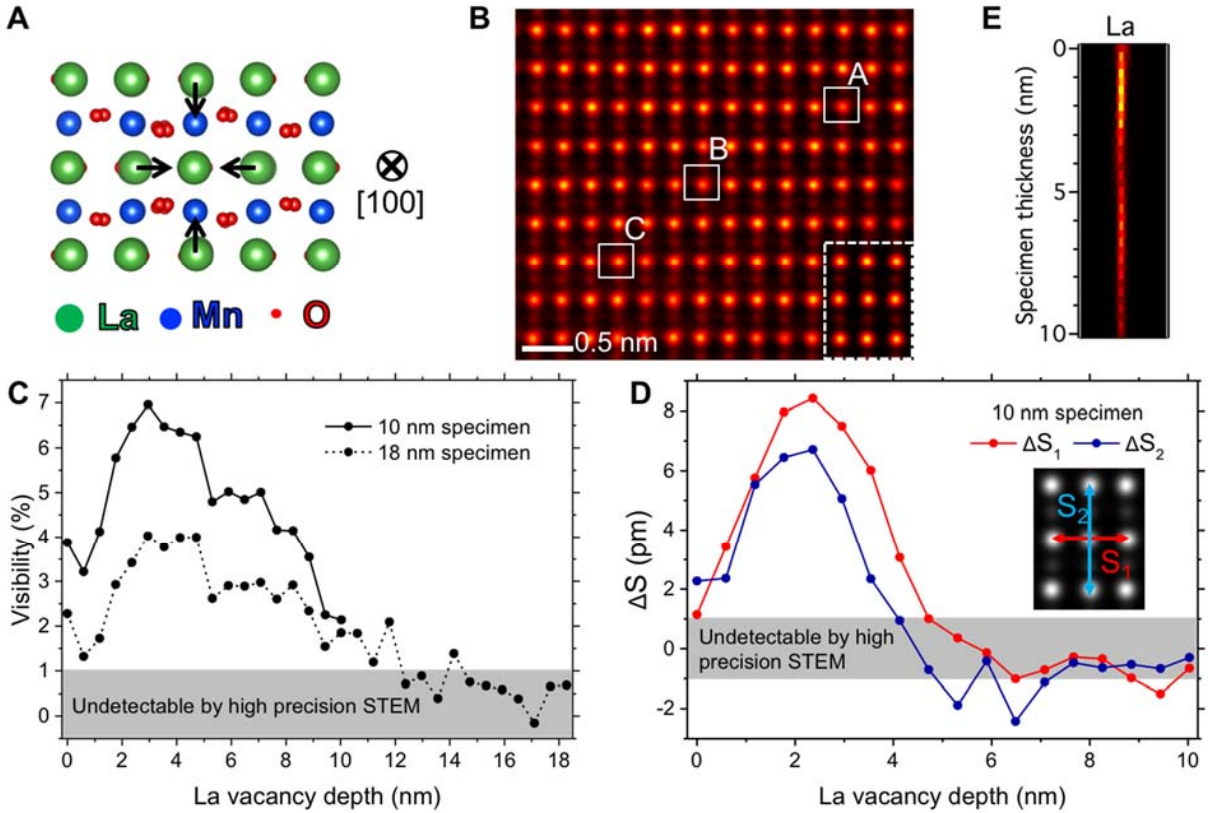

**Fig. 1 HAADF STEM image of LaMnO$_3$ and simulations of visibility and atomic column displacements.** (A) Relaxed crystal structure of LaMnO$_3$ by DFT. (B) Experimental HAADF STEM image on a 6.5 nm thick specimen along [100], with matching simulated image (white box) and projected LaMnO$_3$ unit cell. Displacements of the La-O columns due to a single La vacancy in the center are shown schematically by white arrows. (C) Simulated visibility of a single La vacancy in a 10 nm (solid) and 18 nm (dash) thick LaMnO$_3$ specimen. (D) Simulated $\Delta S$ caused by a single La vacancy in a 10 nm thick specimen. Definitions of $S_1$ and $S_2$ are shown in the inserted HAADF image. (E) Simulated intensity of the electron probe as it propagates along a La-O [100] column. Yellow is higher intensity.

> 1%) with high-precision STEM at all depths. In an 18 nm thick specimen (dash), the vacancy is detectable for depths less than 12 nm. Fig. 1D shows the simulated $\Delta S$'s caused by a La vacancy in a 10 nm specimen. It demonstrates that the $\Delta S$'s are big enough (> 1 pm) to be detected when a La vacancy is within 4 nm of the top surface. Even though the 3D atomic displacements are larger than 20 pm in Fig. 1A, only a small fraction of the displacement is preserved in a HAADF STEM image. Both $V$ and $\Delta S$'s are dependent on the La vacancy depth, which can be explained by the $e^-$ channeling effect shown in Fig. 1E. It shows that when an $e^-$ probe is located on top of a La-O column, the $e^-$ intensity along this column oscillates, and the maximum intensity is at 2 to 3 nm from the entrance surface.

Fig. 2 shows more detailed analysis of the three columns in Fig. 1B labeled A, B, and C. For each column, we compare the experimental measured $V^e$, $\Delta S_1^e$ and $\Delta S_2^e$ to simulations as a function of La vacancy depth drawn from a library of simulations computed as a function of depth and thickness. For column A, $V^e$ = 10.6%, $\Delta S_1^e$ = 3.9 pm, and $\Delta S_2^e$ = 3.0 pm, respectively, all of which are consistent with a vacancy in the column and inconsistent with a vacancy-free column. If we consider a conservative uncertainty band twice the estimated experimental intensity uncertainty of 1%, the visibility is consistent with a vacancy anywhere from 1.1 to 5.2 nm deep in the sample, as shown in Fig 2A. $\Delta S_1^e$ and $\Delta S_2^e$, given an uncertainty of 1 pm, indicate the vacancy is either 0 – 0.8 nm deep or near 3.5 nm deep, as also shown in the figure. The overlap of all three factors localizes the vacancy to 3.4 to 4.2 nm (either 7 or 8 atomic layers) deep into the sample. Fig. 2B shows the same analysis for column B, with $V^e$ = 5.4%, $\Delta S_1^e$ = 3.8 pm, and $\Delta S_2^e$ = 1.4 pm. $V^e$ suggests the vacancy is near the top or the bottom surface of the sample. However, $\Delta S_1^e$ and $\Delta S_2^e$ suggest the top surface or 3.3 – 4.2 nm deep. Considering both factors, the vacancy localizes the vacancy near the top surface.

Fig. 2 A and B show the importance of considering $\Delta S$s in addition to $V$ in the vacancy depth determination: the different depth dependencies of $V$, $\Delta S_1$ and $\Delta S_2$ enable us to distinguish

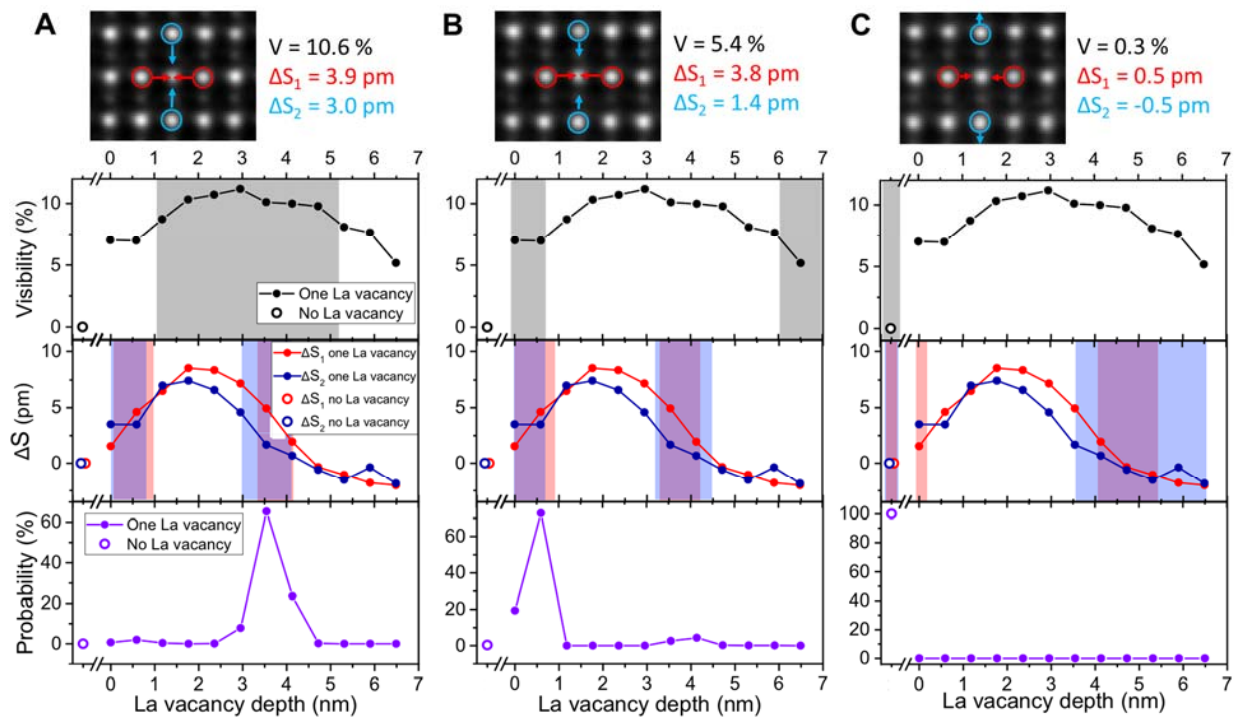

**Fig. 2 Determination of La vacancy lateral position and depth from experimental HAADF images.** (A) – (C) Analysis of atomic columns A, B and C marked in Fig. 1B. From top to bottom: Experimental image and the measured $V$, $\Delta S_1$ and $\Delta S_2$ for each column; simulated visibility and simulated $\Delta S_1$ and $\Delta S_2$ for a TEM specimen of the same thickness; and the Bayesian model probability for the vacancy position. The shaded bands on the simulation results show the range of possible vacancy depths consistent with the experimental visibility and $\Delta S$'s, with a confidence interval of twice the estimated uncertainty (1% for visibility, 1 pm for $\Delta S$). The first data point (open symbol) in the Bayesian model results is the probability of no vacancy in the atomic column.

vacancies inside the thinned TEM specimen, which were present before thinning and would influence the bulk film properties in other measurements, from vacancies on the surface of the

TEM specimen, which were likely introduced by sample thinning. Furthermore, considering both factors overcomes the constraint discussed in previous work on identifying the depth of high atomic number substitutional impurities from the intensity alone, which is that the specimen thickness has to be smaller than half the electron intensity oscillation period along the column[4,5]. For aberration-corrected STEM experiments on high atomic number samples, the oscillation period can be a little as 3 nm (see Fig 1E), making it very difficult to make samples for intensity-only techniques[11]. Our method makes it possible to use thicker specimen and aberration-corrected microscopes, and to detect vacancies at larger depth (see Supplementary Information Note 1 and Fig. S2).

Fig. 2C shows a column that does not contain a vacancy. It has $V^e$ = 0.3%, $\Delta S_1^e$ = 0.5 pm, and $\Delta S_2^e$ = -0.5 pm. $V^e$ is inconsistent with a vacancy at any depth, and $\Delta S_1^e$ and $\Delta S_2^e$ are consistent with no vacancy or a vacancy near the bottom surface, where the column shifts it imposes are nearly zero.

**Bayesian statistical model**

We have developed a Bayesian statistical model to make the graphical analysis of Fig. 2 quantitative. Given a set of measurements, $\boldsymbol{M}^e = (V^e, \Delta S_1^e, \Delta S_2^e)$, on a particular La-O column in an experimental image where the thickness is $t$, the probability that it contains one single La vacancy at the $i^{th}$ atomic layer ($d_i$) can be calculated as:

$$P(d_i|\boldsymbol{M}^e) = \frac{P(V^e|d_i) \times P(\Delta S_1^e|d_i) \times P(\Delta S_2^e|d_i) \times P(d_i)}{P(\boldsymbol{M}^e)}, \qquad (3)$$

$P(V^e|d_i)$, $P(\Delta S_1^e|d_i)$ and $P(\Delta S_2^e|d_i)$ are the probability $V^e$, $\Delta S_1^e$ and $\Delta S_2^e$ are measured in an experiment, respectively, given that a single La vacancy exists in the $i^{th}$ atomic layer in a column. $t$ is decided by PACBED measurement[30]. In Bayes' rule, they are likelihoods, and they can be evaluated from the library of simulations, assuming appropriate uncertainties. $P(d_i)$ and $P(M^e)$ are the prior probability and normalization factor in Bayes' rule, respectively. We assume $P(d_i)$ is constant, and each La-O column contains at most one vacancy, so ( $\sum_i P(d_i|M^e) + P(no\ vac.|M^e) = 1$). This assumption is reasonable due to the small La vacancy concentration in the film and the small thickness of our TEM specimens. The model could extend to two or more vacancies in a column by expanding the library of simulations. It could also be extended to treat other point defects or defect complexes, but Mn vacancies are the only other defects which occur at appreciable concentration in LaMnO$_3$.[33,34] See the Supplementary Information Note 1 and Table S1 for details on evaluating Eq. 3.

The bottom panels of Fig 2 show the result of the Bayesian model for each analyzed column. The probability that a single La vacancy exists in column A is greater than 99.9%, and the most probable depth (65% probability) is 3.5 nm (7$^{th}$ atomic layer). Column B is greater than 99.9% likely to contain a vacancy, and the most probable depth in the first atomic layer, 0.6 nm, with a probability of 73%. Column C is >99.9% likely to contain no vacancy at all.

**Discussion**

The Bayesian model can be applied automatically to a large number of columns to characterize the distribution and clustering of vacancies. However, certain criteria must be met for the model to apply to a La-O column: (A) The intensity fluctuations of the eight surrounding columns, used as $I_{ref}$ in Eq. 1, must be smaller than half of the maximum visibility of a single vacancy at the same thickness, so that the visibility can be reliably calculated. This criterion excludes parts of the STEM sample with local surface roughness that is too high. (B) Each La-O column should be in a low stress area where the spatial measurement precisions are smaller than half of the maximum $\Delta S_1^s$ and $\Delta S_2^s$ of a single La vacancy at the same thickness. This excludes regions of the sample within the strain fields of grain boundaries (Fig. S5(A)) and regions damaged by the ion beam during sample preparation. (C) $V^e$, $\Delta S_1^e$ and $\Delta S_2^e$ must fall within reasonable ranges set based on simulated values for single vacancies. This excludes regions that could contain Mn vacancies or multiple cation vacancies in a column that are not considered in the Bayesian model (see Supplementary Information Note 3 for details). None of these criteria depends on the La vacancy concentration or vacancy position, so it should not influence the vacancy concentration measured using this method. The Bayesian model could be extended with a larger library of simulations to include other types of defects, defect clusters, and uniform strain fields, but the current model focuses only on single vacancies.

Fig. 3A shows all the columns from Fig. 1B that satisfy the criteria above. The number on each column is the Bayesian model probability that it contains one single La vacancy at any depth. The initial La vacancy concentration that determines $P(d_i)$ is assumed to be 1%, and we count a single La vacancy if the Bayes model probability is larger than 95%. However, the results are insensitive to these parameter choices (see Supplementary Information Note 1 and Fig. S1). This analysis was repeated for 20 images, resulting in analysis of 357 La-O columns ranging in thickness from 6 nm to 18 nm. For this range of TEM sample thickness, vacancies are only detectable to a depth of 7 nm. Beyond that depth $V$, $\Delta S_1$ and $\Delta S_2$ are within the experimental uncertainty of zero (Fig. 1 C and D, and Fig. S2). 35 single La vacancies were found in the 357 selected La-O columns. For the column thicker than 7 nm, we studied the 13 atom sites in the top 7 nm, otherwise we studied the whole column. From the total number of possible vacancy sites considered in these columns we measure that the vacancy concentration is 0.79%.

This value is consistent with 1% chemical component control during MBE growth and with the equilibrium concentration of La vacancies under the MBE growth conditions.[33,34]. Point defect concentrations have been studied in bulk LaMnO$_3$ using thermogravimetric analysis[33] and described by defect thermodynamic models [26,33]. Experiments and models agree that only La vacancies and Mn vacancies exist at high PO$_2$ and that the La vacancy concentration (as well as Mn vacancy concentration) is 0.8% under our growth conditions (883 K and 4×10$^{-9}$ atm). MBE is sometimes used to grow non-equilibrium materials, but during our growth, the film was annealed for 30 seconds following the completion of each unit cell to equilibrate it. As shown in Fig. 3B, the 35 single La vacancies are randomly distributed in depth, which is also in consistent with the constant film growth conditions.

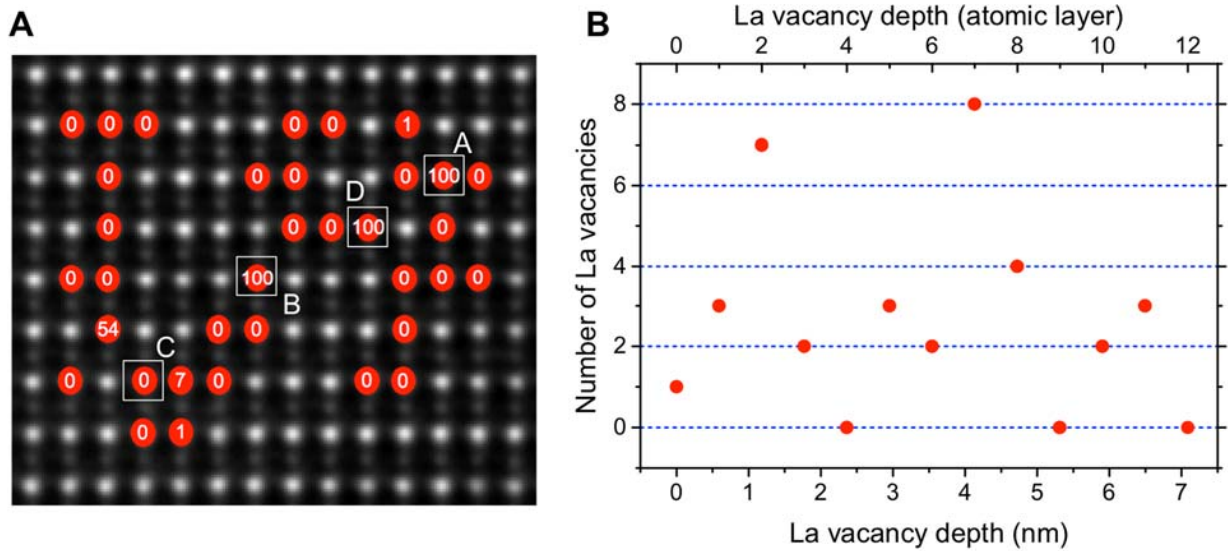

**Fig. 3 Single La vacancy distribution in depth.** (A) La columns in Fig. 1B that pass the criteria for analysis with the Bayesian model. The number is the probability that each column contains one La vacancy calculated by Bayesian model. In this image, three single La vacancies are found in columns A, B and D. (B) Depth distribution of all 35 single La vacancies found in analysis of many images.

In conclusion, we have demonstrated three-dimensional imaging of single La vacancies in bulk LaMnO$_3$ by combining high precision quantitative HAADF STEM, frozen phonon simulation, Bayesian statistical modeling, and DFT. The close match between experimental and simulated HAADF STEM image features, the measured single La vacancy concentration (0.79%), and the random distribution of La vacancies in depth provide confidence that the method is reliable for three-dimensional detection of single vacancies in bulk materials. Extension of the method to vacancies in other materials, detection of vacancy complexes, and treatment of other defects including impurities and even anti site defects is straightforward, and by combining the absolute scattered intensity and lattice distortions, we overcome limitations associated with previous methods for imaging single impurity atoms. Our results represent a step forward in characterizing point defects in materials one at a time, at atomic resolution, matching our current capabilities in materials simulation and our growing control over defect distributions in synthesis. This capability closes the materials design loop from simulation to synthesis to characterization, accelerating our ability to use control over point defects to create new materials functionality and improve performance.

**Experimental Section**

**Density functional theory (DFT) calculations**

DFT calculations were performed with the Vienna Ab-initio Simulation Package (VASP) using a plane wave basis set, the GGA-Perdew-Wang-91 (PW91) exchange-correlation function, and the projector augmented wave (PAW) method. To treat the correlated electron behavior in LaMnO$_3$ in DFT we use the GGA + U method, where $U_{eff} = U - J = 3.5$ eV and is applied to the Mn d orbitals[26]. We used PAW potentials La (5s2 5p6 6s2 5d1), Mn$_{pv}$ (3p6 3d6 4s1) and O$_s$ (soft oxygen potential, 2s2 2p4)[27]. In the DFT calculation, we used a 3 × 2 × 3 supercell of the primitive *Pnma* unit cell, which has a size of 17.7 Å × 15.5 Å × 16.7 Å. The primitive cell has 20 atoms so that the perfect LaMnO$_3$ supercell has (3 × 2 × 3 = 18) × 20 = 360 atoms and the single La vacancy contained supercell has 359 atoms. Such large supercell was used to eliminate the interaction between the La vacancy and its duplicates produced by the period boundary condition in DFT calculation. For the supercell we used a 1 x 1 x 1 Monkhorst-Pack k-point mesh in the Brillouin zone. The ions are relaxed until the Hellmann-Feynman forces are less than 1 meV/Å. Both the perfect supercell and the single La vacancy contained supercell were used to build the LaMnO$_3$ model used in frozen phonon simulation.

**HAADF STEM image simulations**

Frozen phonon simulations as implemented by Kirkland[14] were performed along LaMnO$_3$ [100], using a 33.5 Å × 31.0 Å × 212.3 Å LaMnO$_3$ model which was built from a perfect LaMnO$_3$ supercell and a single La vacancy contained LaMnO$_3$ supercell fully relaxed by DFT (2 × 2 × 12). By changing the depth location of the single La vacancy supercell in the LaMnO$_3$ model, we can simulate the HAADF STEM image with different La vacancy depth.

In frozen phonon simulations, the LaMnO$_3$ model was sampled with a 2048 × 2048 pixel wave function and 128 phonon configurations were used to decrease the intensity standard deviation of each column over ten independent simulations smaller than 1 % and the standard deviation of the column position over ten simulations smaller than 1 pm, which match the precision in experiments. Debeye Waller factors are taken as 0.48 Å$^2$ (La), 0.40 Å$^2$ (Mn), 0.60 Å$^2$ (O1) and 0.40 Å$^2$ (O2) (where O1 and O2 are the two symmetry distinct O in the primitive cell)[35]. The simulated images were convolved with the 86.3 pm full width at half maximum (FWHM) Gaussian function to account for incoherent source broadening[13]. All microscope settings are the same as experiments.

**Gaussian fitting: column intensity and position analysis**

To accurately calculate column intensity and achieve sub-pixel location of the atomic column position, each La-O column was fit to a 2D Gaussian function using the standard Levenberg-Marquart $L^2$-norm minimization method in both experimental and simulated images. The function form for the fitting was

$$z(x,y) = z_0 + A \exp\left[\left(\frac{-1}{2(1-c^2)}\right)\left(\left(\frac{x-x_0}{x_w}\right)^2 + \left(\frac{y-y_0}{y_w}\right)^2 - \left(\frac{2c(x-x_0)(y-y_0)}{x_w y_w}\right)\right)\right], \quad (4)$$

where $z_0$, $A$, $c$, $x_w$, $y_w$, $x_0$ and $y_0$ are fitting parameters. The column position is $(x_0, y_0)$, and intensity is

$$I = z_0 l^2 + 2A\pi |x_w||y_w|\sqrt{1-c^2}, \quad (5)$$

where $l$ is the size of the square fitting region of pixels around each column.

**HAADF STEM experiments**

All high precision quantitative HAADF STEM experiments were performed on a FEI Titan microscope with a CEOS probe aberration-corrector operated at 200 keV where we were able routinely reach 0.8 Å resolution. The probe semi-angle is 24.5 mrad and the probe current is ~25 pA, and all HAADF STEM images were collected by a Fischione Model 3000 detector subtending 84 to 422 mrad. The microscope diffraction lens was adjusted in free lens control to ensure that the whole HAADF detector was illuminated. Each image series was acquired 300 frames using 256 × 256 pixels and the pixel dwell time was ~12 μs.

**LaMnO$_3$ film synthesis**

The sample investigated in this work consists of a 50 unit cell thick LaMnO$_3$ film grown on a DyScO$_3$ (110) single crystal substrate by ozone-assisted molecular beam epitaxy. The film was deposited by coevaporation of La and Mn metals from individual effusion cells, with evaporation rates determined to within 1% by a calibrated quartz crystal thickness monitor. During growth, the substrate was maintained at 610°C under an ozone partial pressure of $2.7 \times 10^{-6}$ mbar. The sample is annealed for 30 seconds under the same conditions following the completion of each unit cell. At the end of growth, the film is allowed to cool to room-temperature with the ozone partial pressure held at $2.7 \times 10^{-6}$ mbar.

The surface crystallinity and film orientation are monitored *in situ* using reflection high-energy electron diffraction (RHEED). The diffraction patterns show that the films are [001]-oriented, with sharp spots lying on a Laue arc, characteristic of atomically smooth surfaces. Oscillations in the intensity of the RHEED diffraction spots indicate that the film growth proceeds in a layer-by-layer fashion.

**STEM sample preparation**

The LaMnO$_3$ samples were mechanically wedge polished at angle 1.6° using Allied MultiPrep Polishing System successively decreasing diamond particle size starting with 15 μm and finishing with finest 0.1 μm lapping diamond film. Then the samples were ion milled from both sides at angle 5° in a Fischione model 1050 TEM mill using Ar gas, starting with 3 kV in the beginning of ion milling, then successively decrease energy of Ar$^+$ using minimal voltage of 100 V. The last step is very important to reduce thickness of damaged/amorphous layer.

**References**

[1]  D. Lee, H. Lu, Y. Gu, S. Ryu, T. R. Paudel, K. Song, E. Mikheev, S. Lee, S. Stemmer, D. A. Tenne, S. H. Oh, E. Y. Tsymbal, X. Wu, A. Gruverman, C. B. Eom, *Science (80-. ).*

**2015**, *439*, 1314.

[2]     M. J. Gadre, Y.-L. Lee, D. Morgan, *Phys. Chem. Chem. Phys.* **2012**, *14*, 2606.

[3]     P. M. Voyles, D. a Muller, J. L. Grazul, P. H. Citrin, H.-J. L. Gossmann, *Nature* **2002**, *416*, 826.

[4]     J. Hwang, J. Zhang, A. D'Alfonso, L. Allen, S. Stemmer, *Phys. Rev. Lett.* **2013**, *111*, 266101.

[5]     R. Ishikawa, A. R. Lupini, S. D. Findlay, T. Taniguchi, S. J. Pennycook, *Nano Lett.* **2014**, *14*, 1903.

[6]     S. H. Oh, K. Van Benthem, S. I. Molina, A. Y. Borisevich, W. Luo, P. Werner, N. D. Zakharov, D. Kumar, S. T. Pantelides, S. J. Pennycook, *Nano Lett.* **2008**, *8*, 1016.

[7]     D. Alloyeau, B. Freitag, S. Dag, L. Wang, C. Kisielowski, *Phys. Rev. B* **2009**, *80*, 14114.

[8]     P. Voyles, D. Chadi, P. Citrin, D. Muller, J. Grazul, P. Northrup, H.-J. Gossmann, *Phys. Rev. Lett.* **2003**, *91*, 125505.

[9]     R. Ishikawa, R. Mishra, A. R. Lupini, S. D. Findlay, T. Taniguchi, S. T. Pantelides, S. J. Pennycook, *Phys. Rev. Lett.* **2014**, *113*, 155501.

[10]    J. Y. Zhang, J. Hwang, B. J. Isaac, S. Stemmer, *Sci. Rep.* **2015**, *5*, 12419.

[11]    H. Kim, J. Y. Zhang, S. Raghavan, S. Stemmer, *Phys. Rev. X* **2016**, *6*, 41063.

[12]    N. Tsvetkov, Q. Lu, L. Sun, E. J. Crumlin, B. Yildiz, *Nat. Mater.* **2016**, 1.

[13]    A. B. Yankovich, B. Berkels, W. Dahmen, P. Binev, S. I. Sanchez, S. a Bradley, A. Li, I. Szlufarska, P. M. Voyles, *Nat. Commun.* **2014**, *5*, 4155.

[14]    Earl J. Kirkland, *Advanced Computing in Electron Microscopy*, Springer, **2010**.

[15] J. LeBeau, S. Findlay, L. Allen, S. Stemmer, *Phys. Rev. Lett.* **2008**, *100*, 206101.

[16] D. a Muller, N. Nakagawa, A. Ohtomo, J. L. Grazul, H. Y. Hwang, *Nature* **2004**, *430*, 657.

[17] Y.-M. Kim, J. He, M. D. Biegalski, H. Ambaye, V. Lauter, H. M. Christen, S. T. Pantelides, S. J. Pennycook, S. V Kalinin, A. Y. Borisevich, *Nat. Mater.* **2012**, *11*, 888.

[18] C. L. Jia, K. Urban, *Science* **2004**, *303*, 2001.

[19] T. Epicier, *J. Am. Ceram. Soc* **1991**, *74*, 2359.

[20] M. Ohwada, K. Kimoto, T. Mizoguchi, Y. Ebina, T. Sasaki, *Sci. Rep.* **2013**, *3*, 2801.

[21] A. Hashimoto, K. Suenaga, A. Gloter, K. Urita, S. Iijima, *Nature* **2004**, *430*, 870.

[22] W. Zhou, X. Zou, S. Najmaei, Z. Liu, Y. Shi, J. Kong, J. Lou, P. M. Ajayan, B. I. Yakobson, J. C. Idrobo, *Nano Lett.* **2013**, *13*, 2615.

[23] M. Choi, F. Oba, I. Tanaka, *Phys. Rev. Lett.* **2009**, *103*, 1.

[24] L. Wang, T. Maxisch, G. Ceder, *Phys. Rev. B - Condens. Matter Mater. Phys.* **2006**, *73*, 1.

[25] J.-S. Zhou, J. Goodenough, *Phys. Rev. B* **1999**, *60*, R15002.

[26] Y.-L. Lee, D. Morgan, *Phys. Chem. Chem. Phys.* **2012**, *14*, 290.

[27] Y.-L. Lee, J. Kleis, J. Rossmeisl, D. Morgan, *Phys. Rev. B* **2009**, *80*, 224101.

[28] M. Choi, F. Oba, Y. Kumagai, I. Tanaka, *Adv. Mater.* **2013**, *25*, 86.

[29] T. N. Tarasenko, a. S. Mazur, O. F. Demidenko, G. I. Makovetskii, K. I. Yanushkevich, *Inorg. Mater.* **2012**, *48*, 1039.

[30] J. M. Lebeau, S. D. Findlay, L. J. Allen, S. Stemmer, *Ultramicroscopy* **2010**, *110*, 118.

[31]  H. E, K. E. MacArthur, T. J. Pennycook, E. Okunishi, a. J. D'Alfonso, N. R. Lugg, L. J. Allen, P. D. Nellist, *Ultramicroscopy* **2013**, *133*, 109.

[32]  J. Feng, A. V Kvit, A. B. Yankovich, C. Zhang, D. Morgan, M. Paul, **2015**, *21*, 1887.

[33]  J. Mizusaki, N. Mori, H. Takai, Y. Yonemura, H. Minamiue, H. Tagawa, M. Dokiya, H. Inaba, K. Naraya, T. Sasamoto, T. Hashimoto, *Solid State Ionics* **2000**, *129*, 163.

[34]  Y. Lee, D. Morgan, **2012**, *14*, 290.

[35]  T. Mori, K. Inoue, N. Kamegashira, *J. Alloys Compd.* **2000**, *308*, 87.

[36]  P. M. Lee, *Bayesian Statistics: An Introduction*, Wiley, **2012**.

[37]  H. Jiang, L. He, D. Morgan, P. M. Voyles, I. Szlufarska, *Phys. Rev. B - Condens. Matter Mater. Phys.* **2016**, *94*, 1.

[38]  R. a. De Souza, M. S. Islam, E. Ivers-Tiffée, *J. Mater. Chem.* **1999**, *9*, 1621.

[39]  N. Alem, O. V. Yazyev, C. Kisielowski, P. Denes, U. Dahmen, P. Hartel, M. Haider, M. Bischoff, B. Jiang, S. G. Louie, A. Zettl, *Phys. Rev. Lett.* **2011**, *106*, 1.

**Acknowledgements**

Authors at UW-Madison acknowledge the funding from the Department of Energy, Basic Energy Sciences (DE-FG02-08ER46547) in support of microscopy experiments, simulations and analysis. UW-Madison facilities and instrumentation are supported by the UW Materials Research Science and Engineering Center (DMR-1121288). We thank Dr. A. Yankovich for the discussion on the STEM image simulation. We thank S. Wang, from Department of Statistics, UW-Madison, for the discussion on statistics. Synthesis of LMO thin films by JH and AB was supported by the Department of Energy, Office of Science, Basic Energy Sciences, Materials Sciences and Engineering Division. The use of facilities at the Center for Nanoscale Materials, an Office of Science user facility, was supported by the U.S. Department of Energy, Basic Energy Sciences under Contract No. DE-AC02-06CH11357.

**Author contributions**

P.M.V and D.M. conceived the study of 3D imaging of single La vacancies by STEM. J.F. performed the DFT calculations, frozen phonon simulations, STEM experiments and data analyses. A.K. and J.F. prepared the STEM experiment samples. J.F., P.M.V. and D.M. conceived the Bayesian statistical model. J.H. and A.B. synthesized the $LaMnO_3$ thin film. J.F. drafted the manuscript, and all authors discussed and revised it.

**Supplemental Materials**

**Supplementary Materials and Methods**

1. Bayesian model

**1.1. Bayesian model building**

The Bayesian statistical model analysis method is built on Bayes' Theorem, which provides a mathematical rule explaining how existing belief should be changed in the light of new evidence[36]. In our application, we start with the assumption that vacancies are uniformly distributed at constant concentration, and then update it to reflect the experimental measured atomic column visibility, $V^e$, and the changes in neighboring column separation, $\Delta S_1^e$ and $\Delta S_2^e$ (the set of measurements on the La-O column in an experimental image is denoted $\boldsymbol{M^e} = (V^e, \Delta S_1^e, \Delta S_2^e)$). The result is the probability that a La-O column contains one single La vacancy at the $i^{th}$ atomic layer ($d_i$), given the experimental data, $P(d_i|\boldsymbol{M^e})$. From Bayes' theorem,

$$P(d_i|\boldsymbol{M^e}) = \frac{P(\boldsymbol{M^e}|d_i) \times P(d_i)}{P(\boldsymbol{M^e})}, \qquad (6)$$

where $t$ is the TEM specimen thickness and decided by PACBED measurement[30]. $P(d_i)$ is the prior probability of a vacancy occurring in the $i^{th}$ atomic layer. $P(\boldsymbol{M^e}|d_i)$ is the likelihood that, if we know a vacancy $d_i$ exists, we obtain measurements $\boldsymbol{M^e}$, given the uncertainties in our experiments and simulations, and $P(\boldsymbol{M^e})$ is a normalization factor.

Implicit in Eq. 6 is that we also include the case in which no vacancy exists to obtain $P(no\ vac.|\boldsymbol{M^e})$. In other words, we allow a special case where $d_i$ represents no vacancy. Therefore, in the Bayesian model, each La-O column is considered either perfect or containing only one single La vacancy. The limit to one vacancy is a reasonable assumption due to the very low La vacancy concentration and small specimen thickness. For example, in a 12 nm thick specimen, the probability that there is more than one La vacancy is smaller than 2%. See Table S1 for details. Instead of calculating $P(\boldsymbol{M}^e)$ explicitly, we ensure that the final probability is normalized such that

$$\sum_i P(d_i|\boldsymbol{M}^e) + P(no\ vac.|\boldsymbol{M}^e) = 1. \tag{7}$$

### 1.1.1 Prior probability $P(d_i)$

$P(d_i)$ is the prior which measures the probability that a single La vacancy existing at the $i^{th}$ atomic layer before any experimental evidence is taken into account. We start with assumption that La vacancies are randomly distributed on the specimen lattice sites with average concentration $c$. As a result, for a column with thickness $t$ containing $n$ La atom sites, the prior probability it contains $s$ La vacancies is given by the binomial distribution,

$$P_{n,c}(s) = \frac{n!}{s!\,(n-s)!} c^s (1-c)^{n-s}. \tag{8}$$

The specimens we used in experiments are stoichiometric LaMnO3 films were grown on DyScO3 substrate by molecular beam epitaxy (MBE) with approximately 1% chemical component control, so 1% is a reasonable starting point for $c$. Furthermore, as shown below in Fig. S1, the results of the Bayesian model are relatively insensitive to the value of $c$. The prior probability that a column is perfect is given by

$$P_{n,c}(0) = (1-c)^n, \tag{9}$$

and $P(d_i)$ for the column which contains one single La vacancy is

$$P(d_i) = \frac{1}{n} P_{n,c}(1) = c^1 (1-c)^{n-1}. \tag{10}$$

### 1.1.2 Likelihood $P(\boldsymbol{M}^e|d_i)$

$P(\boldsymbol{M}^e|d_i)$ is the likelihood that, given a vacancy $d_i$, we obtain a set of measurements in a small range between $\boldsymbol{M}^e$ and $\boldsymbol{M}^e + d\boldsymbol{M}^e$ in experiment. The measured $V^e$, $\Delta S_1^e$ and $\Delta S_2^e$ are all caused by the La vacancy and are dependent on the vacancy depth ($i$) and specimen thickness ($t$), so $V^e$, $\Delta S_1^e$ and $\Delta S_2^e$ are conditionally independent of one another. As a result,

$$P(\boldsymbol{M}^e|d_i) = P(V^e|d_i) \times P(\Delta S_1^e|d_i) \times P(\Delta S_2^e|d_i) = P(V^e|V^r) \times \qquad (11)$$

$$P(\Delta S_1^e|\Delta S_1^r) \times P(\Delta S_2^e|\Delta S_2^r).$$

Here $P(V^e|d_i)$ is the likelihood that we obtain the value experimentally visibility between $V^e$ and $V^e + dV^e$ given a vacancy $d_i$, and $(\Delta S_1^e|d_i)$ and $P(\Delta S_2^e|d_i)$ are defined similarly. $V^r$, $\Delta S_1^r$ and $\Delta S_2^r$ are the "real" noiseless data we should measure with a perfect instrument given a vacancy $d_i$, so we can write $P(V^e|d_i)$ as $P(V^e|V^r)$, which is a clearer notation for the derivation below. All three terms on the right hand side of Eq. 11 can be calculated based on a library of STEM image simulations and incorporating uncertainties in both simulations and experiments. Here we describe the calculations of $P(V^e|V^r)$ in detail. $P(\Delta S_1^e|\Delta S_1^r)$ and $P(\Delta S_2^e|\Delta S_2^r)$ are calculated by an analogous approach.

We treat the experiment HAADF STEM visibility, $V^e$, in the following way:

$$V^e = V^r + \varepsilon. \qquad (12)$$

$V^r$ is the "real" noiseless visibility and $\varepsilon$ is the experimental uncertainty. We assume $\varepsilon$ is given by a normal distribution $N_\varepsilon$ with mean 0 and variance $(\sigma^e)^2$, i.e.,

$$\varepsilon \sim N_\varepsilon(0, (\sigma^e)^2). \qquad (13)$$

Here the tilde signifies that $\varepsilon$ is a random variable with the distribution $N_\varepsilon$. $V^r$ can be estimated from $V^s$, which is the simulated visibility by frozen phonon multislice simulation. The stochastic averaging over phonon displacement configurations means that the simulation also has some uncertainty, $\sigma^s$. We assume that our simulations are exact except for this stochastic uncertainty. With this assumption $V^r$ is dominated by a normal distribution with mean $V^s$ and variance $(\sigma^s)^2$.

$$V^r \sim N_r(V^s, (\sigma^s)^2). \tag{14}$$

Because experiments and simulations are two separated processes, $N_\varepsilon$ and $N_t$ are two independent normal distributions. As a result, the distribution of $V^e$ can be calculated as the sum of $N_\varepsilon$ and $N_t$.

$$V^e \sim N_e(V^s, (\sigma^s)^2 + (\sigma^e)^2). \tag{15}$$

Therefore, given a vacancy $d_i$, the probability we get a visibility measurement in the range of $V^e$ to $V^e + dV^e$ ($P(V^e|V^t)$) can be calculated as:

$$P(V^e|V^t) = N_e(V^s, (\sigma^s)^2 + (\sigma^e)^2) \times dV = \frac{dV}{\sqrt{2\pi}\sqrt{(\sigma^s)^2+(\sigma^e)^2}} e^{-\frac{(V^e-V^s)^2}{2((\sigma^s)^2+(\sigma^e)^2)}}. \tag{16}$$

$\sigma^e$ is estimated at 1% from the counting statistics of the HAADF STEM experiments. The frozen phonon simulations were converged to the match the experimental uncertainty, so $\sigma^s$ is also 1%. For the calculation of $P(\Delta S_1^e|\Delta S_1^r)$ and $P(\Delta S_2^e|\Delta S_2^r)$, $\sigma^e$ is 1 pm based on the precision of previous experiments using high precision STEM experiments[13], and the frozen phonon simulation tests were converged to make $\sigma^s$ also 1 pm.

### 1.2 Bayesian model test

#### 1.2.1 Initial La vacancy concentration ($c_v^{initial}$) and threshold test

As discussed in Eq. 6 to 11, an initial La vacancy concentration ($c_v^{initial}$) is needed in the Bayesian model calculation. In all discussions in the main text and the phantom tests, we use 1% for $c_v^{initial}$, since our samples are LaMnO$_3$ films grown on DyScO$_3$ substrate by molecular beam epitaxy (MBE) with 1% chemical component control. Fig. S1 shows how the Bayesian model measured experimental La vacancy concentration ($c_v$) changes with different $c_v^{initial}$ and threshold values in a set of iterative tests. Four different $c_v^{initial}$, 0.5%, 1.0%, 1.5% and 2.0%, were used and the new calculated $c_v$ was used as the $c_v^{initial}$ in the next iteration. The iteration continues until the new calculated $c_v$ doesn't change. Two different thresholds, 90% and 95%, were used. We claim a La-O column contains a single La vacancy when the probability calculated by the Bayesian model is larger than the threshold value.

As shown in Fig. S1, the Bayesian model measured experimental $c_v$ converges to the same value after 4 to 6 iterations regardless of $c_v^{initial}$ and threshold values. This demonstrates that the Bayesian model is robust to the choice of $c_v^{initial}$ and threshold values. $c_v^{initial}$ was set to be 1% in all Bayesian model analysis and the threshold values was set to be 95%.

1.2.2 Phantom test

In order to validate the Bayesian model, we have done a series of phantom tests. In each test, fake experimental measurements were created by adding noise to the noiseless simulated $V^s$, $\Delta S_1^s$ and $\Delta S_2^s$ from a specific LaMnO$_3$ model. Then we tested whether the Bayesian model correctly recovered whether or not a La vacancy was present and the depth of the vacancy if it was present. The noise was drawn from a normal distributions with mean value of 0, standard deviation of 1% for $V^s$, and 1 pm for $\Delta S_1^s$ and $\Delta S_2^s$. 12 nm thick LaMnO$_3$ models were used in the phantom tests since the average experimental specimen thickness is 12 nm. Each test was repeated with 10000 different noise realizations.

Fig. S2 (A) is a contour plot of the phantom test results with the real vacancy depth in the model on the $x$ axis and the Bayes model result on the $y$ axis. For example, the Bayesian model predicted that there is no La vacancy in the model in all 10000 tests from the LaMnO$_3$ model without a vacancy. For a single vacancy LaMnO$_3$ model with the vacancy at 2.4 nm depth, the

Bayes model found the vacancy at the correct depth in 61% tests, 38% tests had a 1-layer prediction error and only <1% test had a 2-layer error. The threshold value was set to be 95% and $c_v^{initial}$ was set to be 1%. When only ±1 layer prediction uncertainty is allowed, the precision is larger than 85% in most case, which demonstrates the single La vacancy depth calculated by the Bayesian model is very reliable and the experimental La vacancy depth distribution shown in Fig. 3 (B) is of high precision.

Fig. S2 (B) shows the sensitivity of the Bayesian model for detecting single La vacancies the TEM specimen thickness is 12 nm. Sensitivity is defined as following:

$$Sensitivity = \frac{TP}{TP + FN} \times 100\%, \quad (17)$$

TP is true positive (which refers to measuring a vacancy when a vacancy is present) and FN is false negative (which refers to measuring no vacancy when a vacancy is present). The sensitivity in Eq. (17) measures the fraction of vacancies that are detected by the Bayesian model among all vacancies in the system. When the La vacancy depth is 1 nm to 4 nm, the sensitivity is above 90%, which indicates the Bayesian model has a high probability (> 90%) correctly finding a single La vacancy and accurately calculating its depth is this situation. When the vacancy is very close to the top surface or its depth is 5 nm to 7 nm, the sensitivity is around 50%, which means, under such conditions, there is a large chance that the Bayesian model fails to detect the existing single La vacancy. When the La vacancy depth is larger than 7 nm, the sensitivity drops quickly to very close to zero, which indicates that the Bayesian model is only robustly able to detect single La vacancies whose depth is smaller than 7 nm.

Fig. S2 (C) is the precision of the Bayesian model in the phantom tests and the precision is defined as:

$$Precision = \frac{TP}{TP + FP} \times 100\%, \tag{18}$$

where TP is true positive, as defined above, and FP is false positive (which refers to measuring a vacancy when a vacancy is not present). Precision gives the fraction of correct predictions among all positive predictions made by the model. As shown in Fig. S2 (C), when the La vacancy depth is smaller than 7 nm, the precision is 100%, which indicates that as long as the model detects a La vacancy within 7 nm, there is always a La vacancy in the column (red). The precisions when the La vacancy depth is larger than 7 nm are not calculated because few or no La vacancies are predicted in this range in the phantom tests..

Figure S3 is the receiver operator characteristic (ROC) curve of the Bayesian model drawn from the phantom tests data. It shows how the model true positive rate changes as a function of false positive rate for changing threshold values, and thereby illustrates the performance of the Bayesian model as the threshold value changes. As shown in Sig. S3, the model true positive rate increases quickly to 0.95 when the false positive rate is still only about 0.05. This indicates that the Bayesian model is very robust to the choice of threshold value and rarely predicts a single La vacancy when the vacancy is not present.

2. **STEM sample thickness measurement**

The experimental sample thickness was measured by position average converged beam electron diffraction (PACBED)[30]. An example is shown in Fig.S4, where by comparing simulated and experimental PACBED patterns, the sample thickness was decided to be 6.5 nm. The comparison is based on the pattern of intensity inside the $\langle 0 \rangle$ disk of the PACBED pattern, which is not detected in the HAADF images, so it provides an independent thickness measurement.

### 3. Column selection on experimental HAADF STEM images

In experiment, we took images on all thin LaMnO$_3$ film areas. However, there were some "bad" areas that damaged by the ion beam during the sample preparation, some areas where the surface is not flat, and some areas are close to the grain boundaries (see Fig. S5 (A)). Such areas should not be analyzed because it is impossible to accurately measure visibility and inter-column separations. Some criteria were set before data analysis, for automatically selection of "good" columns by computer.

#### 3.1 Local flatness

For each La-O column in the experimental images, the eight surrounding La-O columns were used as the intensity reference to calculate its visibility, as shown in Eq. 1 in the main text. To ensure the accuracy of visibility calculation, the intensity fluctuation on the eight columns should satisfy

$$Max(V^s(t)) > 2\frac{\sigma_I}{\bar{I}}, \tag{19}$$

where $V_{sim}(t)$ is the simulated visibility caused by a single La vacancy when the specimen thickness is $t$. $\bar{I}$ and $\sigma_I$ are the mean intensity and the intensity standard deviation of the eight surrounding La-O columns, respectively. This also ensures the studied La-O column is in a locally flat area.

#### 3.2 Low stress

For each La-O column in the experimental image, requirements on its local spacial measurement precisions ($\sigma_{S_1^e}$ and $\sigma_{S_2^e}$) were set as Eq. 20. $\Delta S_1^s(t)$ and $\Delta S_2^s(t)$ are the simulated inter-column separation changes, defined in Fig. 1 (D) caused by a single La vacancy when the specimen thickness is $t$. $\sigma_{S_1^e}$ and $\sigma_{S_2^e}$ are the precisions of $S_1^e$ and $S_2^e$ in a rectangle area containing the studied La-O column. The rectangle area should contain more than 5 La-O columns along $S_1^e$ direction and 3 La-O columns along $S_2^e$ direction. This ensures that, in experimental images, the computer procedure excludes the areas that were damaged in the sample preparation or close to the grain boundaries where the measured $\Delta S_1$ and $\Delta S_2$ in these areas were not caused by single La vacancies.

$$Max(\Delta S_1^s(t)) > 2\sigma_{S_1^e}$$
$$Max(\Delta S_2^s(t)) > 2\sigma_{S_2^e}$$
(20)

### 3.3 Excluding other kinds of point defect

Other point defects, such as single Mn vacancies, may also exist in the thin film samples we have examined. They are not part of the simulation library, so the Bayesian model will be systematically incorrect if applied to images of such defects. Therefore, we apply several criteria are needed on the range of experimental measured $V$, $\Delta S_1$ and $\Delta S_2$ to exclude these other point defects. We have used

$$Min(V^s(t)) - 1 < V^e < Max(V^s(t)) + 1$$
$$Min(\Delta S_1^s(t)) - 1 < \Delta S_1^e < Max(\Delta S_1^s(t)) + 1 \qquad (21)$$
$$Min(\Delta S_2^s(t)) - 1 < \Delta S_2^e < Max(\Delta S_2^s(t)) + 1$$

where $V^e$, $\Delta S_1^e$ and $\Delta S_2^e$ are the experimental measurements on a La-O column and the specimen thickness is $t$. $V^s(t)$, $\Delta S_1^s(t)$ and $\Delta S_2^s(t)$ are the corresponding simulated values.

About 20% of the atomic columns in the STEM images satisfy all three criteria in (16), which is reasonable given the number of columns that must be defect free for a particular column to pass muster. For a La-O column (black box in Fig. S5(B)) to be analyzed, the positions of the four neighboring La-O columns (red box) must be accurately measured. This requires that all columns in the blue boxes are vacancy-free in the top 5-7 layers. Given the estimated La vacancy concentration of 0.79 %, for one column, the probability that top 6 layers are perfect is 95.35% and the probability that all 24 columns have perfect top 6 layers is about 31%. If another 9% of columns are excluded due to surface roughness, we arrive at 20% columns that satisfy all the criteria.

The density of vacancies detected in our experiments is not sensitive to the threshold values in Eq. 21. If we change the first two criteria to the following,

$$Max(V^s(t)) > \frac{\sigma_I}{\bar{I}}$$

$$Max(\Delta S_1^s(t)) > \sigma_{S_1^e}$$

$$Max(\Delta S_2^s(t)) > \sigma_{S_2^e}$$

and repeat the analysis, we find a concentration of 0.83%. We identify 31 more vacancies, but from proportionally more atomic columns. However, the newly identified vacancies tend to occupy columns with either 0 or negative $V$, inconsistent with a vacancy, but with positive $\Delta S_1$ and $\Delta S_2$, consistent with the vacancy.

4. **Stability of La vacancy in LaMnO$_3$ during the experiment**

The energy transferred from an incident electron to an atom in the sample, $E_n^{'}$, is given by,[37]

$$E_n^{'} = \frac{E(E + 1.02)}{496A} sin^2 \frac{\theta}{2} \qquad (22)$$

where θ is the electron scattering angle, A is the atomic mass, and E is the energy of incident electron. E and $E_n^{'}$ have the unit of MeV. From Eq. 22, the maximum energy that can be transferred from 0.2 MeV electron beam to a La atom is 3.54 eV, which is smaller than the activation energies for La vacancy migration, which are in the range of 4.14 – 4.32 eV depending on the different migration paths[38]. The threshold energy of creating a La vacancy is generally larger than the vacancy mediated migration energy, so the incident beam neither creates nor moves La vacancies during the STEM experiment.

## 5. DFT calculations on different vacancy charge state

In some materials with a large band gap and localized charges, the structure of a defect structures can be influenced by its charge state[39]. We have relaxed the structure of La vacancies in various charge states using DFT. These calculations show that the La vacancy structure in LaMnO$_3$ changes only slightly with charge state. The resulting small differences are well within our experimental uncertainty, so they have no effect on the results.

In DFT for a neutral supercell, $\Delta S_1$ = 20.3 pm and $\Delta S_2$ = 21.9 pm projected on the [100] plane. In a fully charged supercell (adding 3 e- in the supercell), $\Delta S_1$ = 22.7 pm and $\Delta S_2$ = 22.6 pm. Based on Fig. 1D in the main text, these changes in $\Delta S_1$ and $\Delta S_2$ correspond to at most a 0.6 pm change in a measured distance in the HAADF STEM images, which is smaller than our experimental uncertainty of 1 pm. Intermediate charge states (1 or 2 e- in the supercell) generate even smaller shifts as shown in Table S2.

**Supplementary Figures**

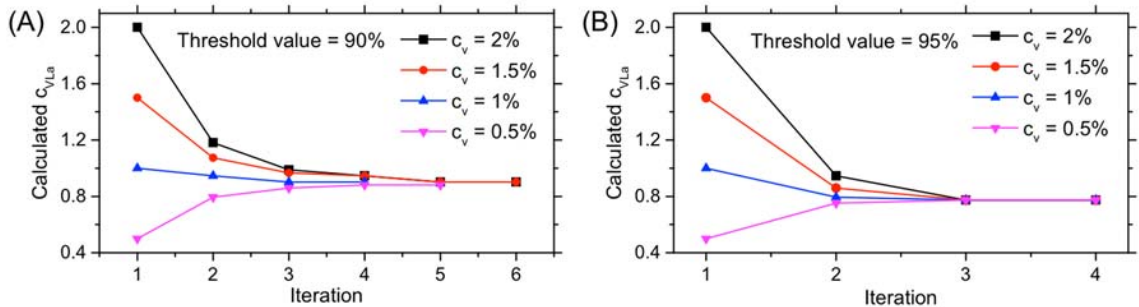

**Fig. S1.** Bayesian model initial La vacancy concentration test.

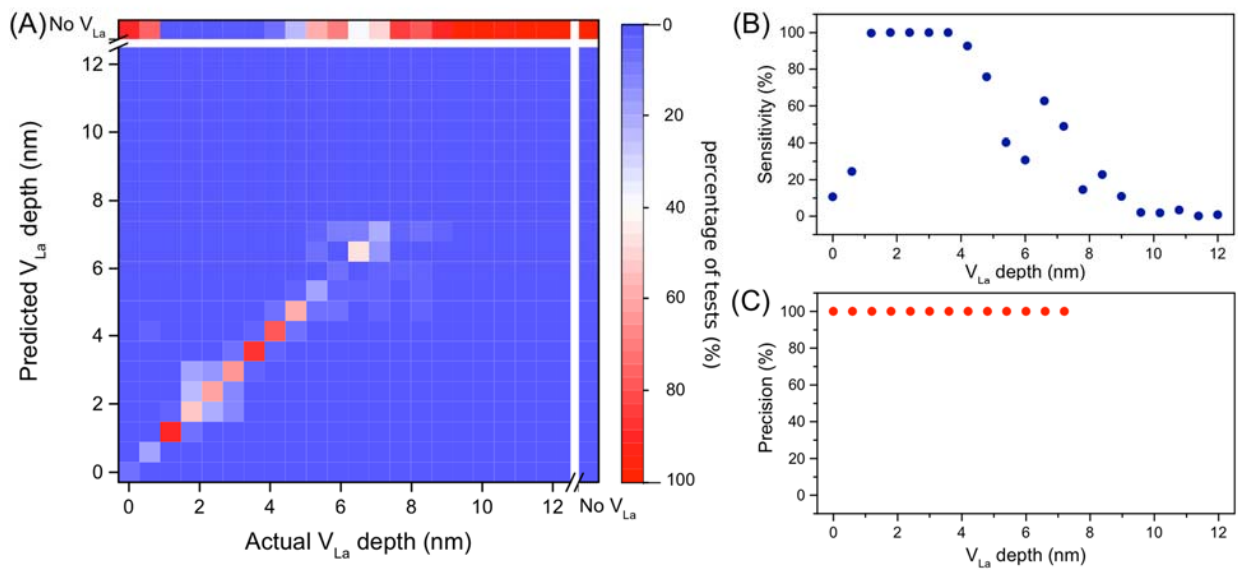

**Fig. S2.** Phantom tests on the Bayesian model. (a) Predicted La vacancy depth *vs.* the actual La vacancy depth. The black line marks the correct prediction. (b) The La vacancy detection sensitivity of the Bayesian model on 18 nm LaMnO3 models. (c) The La vacancy detection precision of the Bayesian model on 18 nm LaMnO3 models.

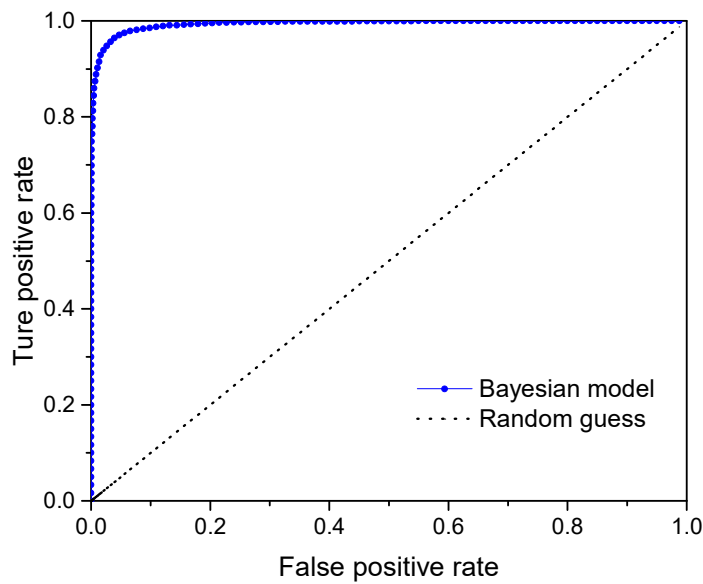

**Fig. S3.** ROC curve of the Bayesian model.

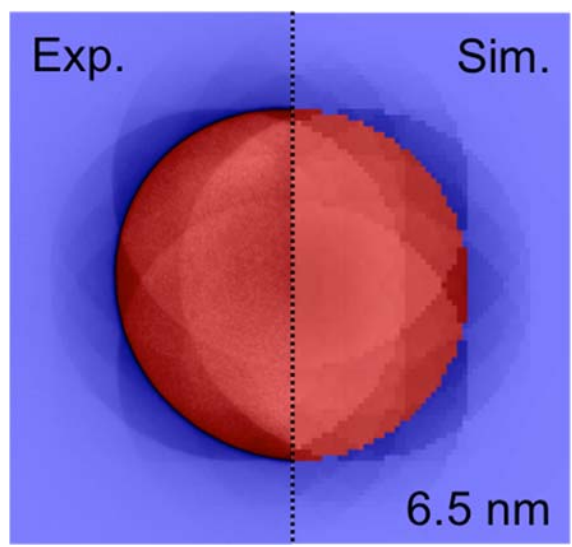

**Fig. S4.** Sample thickness measurement by comparing experimental PACBED pattern (left) with simulated PACBED pattern (right).

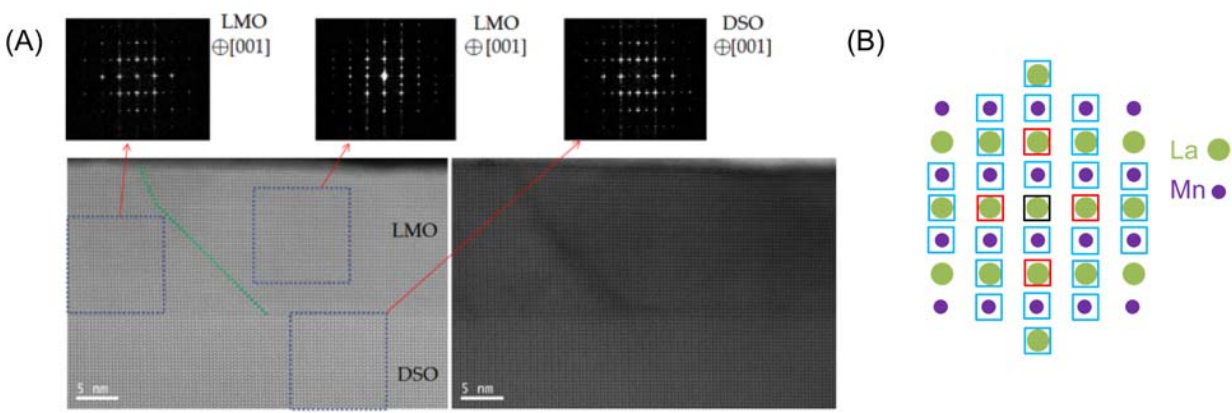

**Fig. S5.** (A) Grains in different orientations in the LaMnO3 film grown on DyScO3 substrate and corresponding FFT. (B) Illustration of the neighborhood surrounding a La-O column (black box) analyzed by the Bayesian vacancy-localization model.

**Supplementary Tables**

**Table S1.** The calculated probability of different number of La vacancy in one La-O column for different specimen thickness when $c$ is 1%.

| Specimen thickness (nm) | number of La site | Prob. of no La vacancy (%) | Prob. of one La vacancy (%) | Prob. of more than one La vacancy (%) |
|---|---|---|---|---|
| 1 | 1 | 99.00 | 1.00 | 0.00 |
| 2 | 3 | 97.03 | 2.94 | 0.03 |
| 3 | 5 | 95.10 | 4.80 | 0.10 |
| 4 | 6 | 94.15 | 5.71 | 0.15 |
| 5 | 8 | 92.27 | 7.46 | 0.27 |
| 6 | 10 | 90.44 | 9.14 | 0.43 |
| 7 | 11 | 89.53 | 9.95 | 0.52 |
| 8 | 13 | 87.75 | 11.52 | 0.72 |
| 9 | 15 | 86.01 | 13.03 | 0.96 |
| 10 | 16 | 85.15 | 13.76 | 1.09 |
| 11 | 18 | 83.45 | 15.17 | 1.38 |
| 12 | 20 | 81.79 | 16.52 | 1.69 |
| 13 | 22 | 80.16 | 17.81 | 2.02 |
| 14 | 23 | 79.36 | 18.44 | 2.20 |
| 15 | 25 | 77.78 | 19.64 | 2.58 |
| 16 | 27 | 76.23 | 20.79 | 2.97 |
| 17 | 28 | 75.47 | 21.35 | 3.18 |
| 18 | 30 | 73.97 | 22.42 | 3.61 |
| 19 | 32 | 72.50 | 23.43 | 4.07 |
| 20 | 33 | 71.77 | 23.92 | 4.30 |

Table S2. $\Delta S_1$ and $\Delta S_2$ in different supercell charge states as calculated in DFT.

| | Neutral supercell | Adding 1 $e^-$ | Adding 2 $e^-$ | Adding 3 $e^-$ |
|---|---|---|---|---|
| $\Delta S_1$ (pm) | 20.3 | 21.2 | 22.0 | 22.7 |
| $\Delta S_2$ (pm) | 21.9 | 22.2 | 22.4 | 22.6 |